\documentstyle[12pt]{article}

\def\be{\begin{equation}}
\def\ee{\end{equation}}
\def\bea{\begin{eqnarray}}
\def\eea{\end{eqnarray}}

\begin{document}

\begin{flushright}
hep-th/0103262
\end{flushright}

\pagestyle{plain}

\def\e{{\rm e}}
\def\haf{{\frac{1}{2}}}
\def\tr{{\rm Tr\;}}
\def\goes{\rightarrow}
\def\gym{g_{{}_{YM}}}
\def\iphi{{\bf i}_\Phi}
\def\iv{{\bf i_v}}
\def\ct{C^{(1)}_t}
\def\ci{C^{(1)}_i}
\def\cj{C^{(1)}_j}
\def\xgphi{{x\goes\Phi}}
\def\ie{{\it i.e.}, }
\def\jbf{{\bf j}}
\def\vbf{{\bf v}}
\def\qbf{{\bf q}}
\def\tcl{T_{\rm cl}}

\begin{center}
\vspace{3cm}
{\Large {\bf On Non-Abelian Structure From Matrix Coordinates}}

\vspace{1cm}

Amir H. Fatollahi 
\footnote{
On leave from: Institute for Advanced Studies in Basic Sciences (IASBS),
Zanjan, Iran.}
${}^,$
\footnote{Postal address: INFN, Sezione di Roma, Dipartimento di Fisica,
Universita di Roma ``La Sapienza", Piazzale Aldo Moro, 2, I-00185 Roma,
Italy.}

\vspace{.5cm}

{\it Dipartimento di Fisica, Universita di Roma ``Tor Vergata",}\\ {\it
INFN-Sezione di Roma II, Via della Ricerca Scientifica, 1,}\\ {\it 00133, Roma,
Italy}

\vspace{.3cm}

{\sl fatho@roma2.infn.it}

\vskip .5 cm
\end{center}

\begin{abstract}
We consider the matrix quantum mechanics of $N$ D0-branes in the
background of the 1-form RR field. It is observed that the transformations
of matrix coordinates of D0-branes induce on the Abelian RR field symmetry
transformations that are like those of non-Abelian gauge fields. The
Lorentz-like equations of motion for matrix coordinates are derived. The
field strengths appearing in the Lorentz-like equations transform in the
adjoint representation of $U(N)$ under symmetry transformations. A
possible relation between D0-brane dynamics in RR background, and the
semi-classical dynamics of charged particles in Yang-Mills background is
mentioned. 
\end{abstract}

\vspace{1.0cm}

\newpage

One of the most interesting aspects of D$p$-brane \cite{9510017} dynamics
is the appearance of ``matrix coordinates" as the dynamical variables
describing the position of coincident D$p$-branes. From String Theory
point of view, this enhancement of degrees of freedom from numbers to
matrices is due to the addition of dynamics of strings stretched between
D$p$-branes to the usual degrees of freedom capturing the dynamics of each
D$p$-brane individually. Consequently, it is understood that the correct
degrees of freedom for a system of bound state of D$p$-branes (and
strings) are matrices \cite{9510135}. 

Though the appearance of matrix coordinates is interesting, a more
interesting case arises when D$p$-branes are put in non-trivial
backgrounds, namely non-trivial form fields. Accordingly, the situation
with matrix coordinates will be serious while it is understood that the
position dependences of background fields on transverse directions of
D$p$-branes also should be given by matrix coordinates. Now a natural
question can be about the consequences of the enhancement of degrees of
freedom on the symmetry issues, and one of the important ones symmetry
transformations. As we will see for the case of D0-branes, the symmetry
transformations of matrix coordinates induce on the 1-form RR field
symmetry transformations which are like those of non-Abelian gauge fields.
In other words, the new transformations of the RR fields are the result of
``active transformations on the matrix coordinates of space." 

We also investigate the covariance of the equations of motion under the
symmetry transformation of the 1-form RR fields. It is observed that the
field strengths appearing in the equations of motion transform under
symmetry transformations as those of Yang-Mills theory, \ie in adjoint
representation.  Finally, we comment on a possible relation between the
dynamics of D0-branes and the semi-classical dynamics of charged particles
in Yang-Mills backgrounds.

To be consistent with T-duality of String Theory, Myers \cite{9910053}
proposed an action containing the Born-Infeld and Chern-Simons parts for
the dynamics of D$p$-brane bound states (see also \cite{9910052}). The
proposed bosonic action for the bound state of $N$ D$p$-branes (in units
$2\pi l_s^2=1$) is the sum of: 
\bea
\!\!\!\!\!\!\!\!\!&~&S_{BI}=-T_p\int d^{p+1}\sigma^a\; \tr\bigg(\e^{-\phi}
\sqrt{-\det(P\{E_{ab}+E_{ai}(Q^{-1}-\delta)^{ij}E_{jb}\}+F_{ab})
\det(Q^i_j)}\bigg),\\
\!\!\!\!\!\!\!\!\!&~&S_{CS}=\mu_p\int\tr\bigg(P\{\e^{i\;\iphi\iphi}(\sum
C^{(n)}\e^{B})\}
\e^{F}\bigg),
\eea
with following definitions:
\bea
&~&E_{\mu\nu}\equiv G_{\mu\nu}+B_{\mu\nu},\;\;\;\;
Q^i_j\equiv \delta^i_j+i[\Phi^i,\Phi^j]E_{kj},\\
&~&\mu,\nu=0,\cdots,9,\;\;\;a,b=0,\cdots,p,\;\;\;i,j=p+1,\cdots,9.\nonumber
\eea
In the above $G_{\mu\nu}$ and $B_{\mu\nu}$ are the metric and the NS-NS
2-form, and $\Phi^i$ are world-volume scalars and $N\times N$ hermitian
matrices, that describe the position of D$p$-branes in the $9-p$ transverse
directions. The $C^{(n)}$ is $n$-form RR field, while $F$ is the $U(N)$
field strength.  In this action, $P\{\cdots\}$ denotes the pull-back of the
bulk fields to the world-volume of the D$p$-branes, and $\tr$is trace on
the gauge group. $\iv$ denotes the interior product with a vector ${\bf
v}$; for example, $\iphi$ acts on a 2-form $C^{(2)}=\haf C^{(2)}_{ij} dx^i
dx^j$ as
\bea
\iphi C^{(2)}= \Phi^i C^{(2)}_{ij} dx^j,\;\;\;\;\;\;
\iphi\iphi C^{(2)}= \Phi^i \Phi^j C^{(2)}_{ij}=\haf [\Phi^i,\Phi^j]
C^{(2)}_{ij}.
\eea
Therefore $(\iphi)^2C^{(n)}$=0 for the commutative case, \ie for one
D$p$-brane. 

Some comments on the above action are in order:

{\it i)} All the derivatives in the longitudinal directions should be
actually covariant derivatives, \ie $\partial_a\goes
D_a=\partial_a+i[A_a,\;]$ \cite{hulldorn}.  This point is true also for the
pull-back quantities.

{\it ii)} The pull-back quantities depend on the transverse directions of
the D$p$-branes only via their functional dependence on the world-volume
scalars $\Phi^i$ \cite{douglas}, ordered by ``symmetrization prescription" 
\cite{9910053,9809100,0010122,9712185,9712159}. For example for the case of
metric $G_{\mu\nu}(x^\rho)$, we can present the $\Phi$ dependences by a
non-Abelian Taylor expansion as \cite{9910053}
\bea
G_{\mu\nu}(\sigma^a,x^i)|_\xgphi&\equiv& 
G_{\mu\nu}(\sigma^a,\Phi^i)=\exp[\Phi^i\partial_{x^i}] 
G_{\mu\nu}(\sigma^a,x^i)\nonumber\\
&=&\sum_{n=0}^\infty
\frac{1}{n!} \Phi^{i_1}\cdots\Phi^{i_n} 
(\partial_{x^{i_1}}\cdots\partial_{x^{i_n}})
G_{\mu\nu}(\sigma^a,x^i)|_{x^i=0}.
\eea

{\it iii)} This action involves a single $\tr$, and this $\tr$should be
calculated by symmetrization prescription for the non-commutative
quantities $F_{ab}$, $D_a\Phi^i$ and $i[\Phi^i, \Phi^j]$ \cite{tseytlin}
\footnote{There is a stronger prescription, with symmetrization between all
non-commutative objects $F_{ab}$, $D_a\Phi^i$, $i[\Phi^i, \Phi^j]$, and the
individual $\Phi$'s appearing in the functional dependences of the
pull-back fields \cite{9910053,9904095}. We do not use this one here, with
no essential change in the conclusions.}. 

Let us consider the special case of D0-branes, with
$G_{\mu\nu}=\eta_{\mu\nu}$ and $B_{\mu\nu}=0$. The low energy action, with
only non-vanishing the RR 1-form, is given by ($\sigma^0=t$) \cite{9910053}
\bea\label{action}
S= \int dt\;\tr \bigg(\haf m_0 D_t\Phi^iD_t\Phi_i 
-\mu_0 \ct(\Phi,t)-\mu_0 D_t\Phi^i \ci(\Phi,t) 
-V(\Phi)\bigg),\\
V(\Phi)=-\frac{1}{4} m_0 
[\Phi^i,\Phi^j]^2,\;\;\;\;\;\;\;D_t=\partial_t+i[A_0(t),\;],
\eea
in which $\ct(\Phi,t)$ and $\ci(\Phi,t)$ are the pull-backs of the RR
1-form $C^{(1)}_\mu(x^\nu)$ to the world-line of the D0-brane bound state. 
Let us first check the effect of a $U(1)$ gauge transformation of the
1-form bulk field $C^{(1)}_\mu(x^\nu)$, defined by
\bea\label{u1-tra}
C^{(1)}_\mu(x^\nu)\goes C'^{(1)}_\mu(x^\nu)=C^{(1)}_\mu(x^\nu)-\partial_\mu
\Lambda(x^\nu),
\eea
with $\Lambda(x^\nu)$ an arbitrary function in the bulk.  Under this
transformation, the variation of the action (\ref{action}) is
\bea
\delta S \sim \mu_0 \int dt\; \tr \bigg(
\partial_t \Lambda(\Phi,t) + 
D_t\Phi^i \partial_i\Lambda(x,t)|_\xgphi \bigg),
\eea
in which the symmetrization prescription is understood after the
replacements $x\goes \Phi$. One then obtains
\bea
\delta S \sim \mu_0 \int dt\; \tr \bigg(
\partial_t \Lambda(\Phi,t) +
\partial_t\Phi^i \partial_i\Lambda(x,t)|_\xgphi 
+i[A_0,\Phi^i]\partial_i\Lambda(x,t)|_\xgphi \bigg).
\eea
The first two terms yield as the surface term $d \Lambda(\Phi,t) /dt$, and
therefore
\bea
\delta S &\sim& \mu_0 \int dt\; \tr \bigg(
i[A_0,\Phi^i]\partial_i\Lambda(x,t)|_\xgphi\bigg)\\ 
\label{vanishing}
&\sim& \mu_0 \int dt\; \tr \bigg(
i A_0[\Phi^i,\partial_i\Lambda(x,t)|_\xgphi]\bigg). 
\eea
At first look this term seems non-vanishing, but in fact due to
symmetrization prescription, it also vanishes \cite{0010122}.  Hence,
thanks to the symmetrization, the action (\ref{action}) is invariant under
$U(1)$ gauge transformations in the bulk. 

Actually the action (\ref{action}) is also invariant under active 
transformations of coordinates, as
\bea\label{phi-tra}
\Phi^i&\goes&\tilde\Phi^i=U^\dagger(\Phi,t)\Phi^i U(\Phi,t)\nonumber\\
A_0(t)&\goes& \tilde A_0(\Phi,t)=U^\dagger(\Phi,t) A_0(t) U(\Phi,t) -i
U^\dagger(\Phi,t)
\partial_t U(\Phi,t),
\eea
with $U(\Phi,t)$ as an arbitrary $N\times N$ unitary matrix;
in fact under these transformations one obtains
\bea\label{DTF}
D_t\Phi^i&\goes&\tilde
D_t\tilde\Phi^i=U^\dagger(\Phi,t)D_t\Phi^iU(\Phi,t),\nonumber\\
D_tD_t\Phi^i&\goes&\tilde D_t \tilde D_t
\tilde\Phi^i=U^\dagger(\Phi,t)D_tD_t\Phi^iU(\Phi,t).
\eea
Now, in the same spirit as for the previously introduced $U(1)$ symmetry of
eq.(\ref{u1-tra}), one finds the symmetry transformations: 
\bea\label{tot-tra}
\Phi^i&\goes&\tilde\Phi^i=U^\dagger(\Phi,t)\Phi^i
U(\Phi,t)\nonumber\\
A_0(t)&\goes&\tilde A_0(\Phi,t)=U^\dagger(\Phi,t) A_0(t) U(\Phi,t) -i
U^\dagger(\Phi,t) \partial_t U(\Phi,t),\nonumber\\
C^{(1)}_i(\Phi,t)&\goes& \tilde C^{(1)}_i(\Phi,t)=
U^\dagger(\Phi,t)C^{(1)}_i(\Phi,t)U(\Phi,t)-iU^\dagger(\Phi,t)
\partial_i U(x,t)|_\xgphi,\nonumber\\
C^{(1)}_t(\Phi,t)&\goes& \tilde C^{(1)}_t(\Phi,t)=
U^\dagger(\Phi,t)C^{(1)}_t(\Phi,t)U(\Phi,t)-iU^\dagger(\Phi,t)
\partial_t U(\Phi,t),
\eea
in which we assume that $U(\Phi,t)=\exp(-i\Lambda)$ is arbitrary up to this
condition that $\Lambda(\Phi,t)$ is totally symmetrized in the $\Phi$'s.
The above transformation on the 1-form RR field is similar to those of
non-Abelian gauge theories, and we see that it is just the consequence of
the existing matrix coordinates. In other words, a $U(1)$ theory on a
matrix coordinate space has symmetry transformations like those of a
non-Abelian theory.

The above observation on gauge theory associated to D0-brane matrix
coordinates on its own is not a new one, and we already know another
example of this kind in non-commutative gauge theories. In spaces whose
coordinates satisfy the algebra
\bea
[x^\alpha,x^\beta]=i\theta^{\alpha\beta}, 
\eea 
with constant $\theta^{\alpha\beta}$, the symmetry transformations of the
$U(1)$ gauge theory are like those of non-Abelian gauge theory, and are
known as non-commutative $U(1)$ transformations \cite{9908142,CDS,jabbari}.
Note that the above algebra is satisfied also for the transformed
coordinates $\tilde{x}^\alpha\equiv U^\dagger(x) x^\alpha U(x)$. 

In addition, the case we see here for D0-branes may be considered as
another example of the relation between gauge symmetry transformations and
transformations of matrix coordinates \cite{0007023}. 

The last notable points are about the behaviour of $A_0(t)$ and
$C^{(1)}_t(\Phi,t)$ under symmetry transformations (\ref{tot-tra}). From
the world-line theory point of view, $A_0(t)$ is a dynamical variable, but
$C^{(1)}_t(\Phi,t)$ should be treated as a part of background, however they
behave similarly under transformations. Also we see by (\ref{tot-tra}) 
that the time, and only time dependence of $A_0(t)$, which is the
consequence of dimensional reduction, should be understood up to a gauge
transformation. 

As expected, the action (\ref{action}) looks like that of an electric
charged particle in an electromagnetic background $(\ct(x,t),\ci(x,t))$. 
Thus in principle, one expects to obtain Lorentz-like equations of motion
from this action. For the moment ignoring the potential term $V(\Phi)$, one
can derive the equations of motion for $\Phi^i$ and $A_0$ as
\bea\label{gle}
\label{lorentz}
m_0D_tD_t\Phi_i&=&\mu_0\bigg(E_i(\Phi,t)+
\underbrace{D_t\Phi^jB_{ji}(\Phi,t)}\bigg)\\
\label{A0}
m_0[\Phi_i,D_t\Phi^i]&=&\mu_0[\ci(\Phi,t),\Phi^i],
\eea
with the definitions:
\bea
\label{elec}
E_i(\Phi,t)&\equiv&-\partial_i\ct(x,t)|_\xgphi+\partial_t\ci(\Phi,t),\\
\label{magn}
B_{ji}(\Phi,t)&\equiv&\partial_j\ci(x,t)|_\xgphi-\partial_i\cj(x,t)|_\xgphi
\eea
where the symbol $\underbrace{D_t\Phi^j B_{ji}(\Phi,t)}$ denotes the
average over all of positions of $D_t\Phi^j$ between the $\Phi$'s of
$B_{ji}(\Phi,t)$. As mentioned, the above equations for the $\Phi$'s are
like the Lorentz equation of motion, with the exceptions that two sides are
$N\times N$ matrices, while the time derivatives $\partial_t$ are replaced
by their covariant counterpart $D_t$ \footnote{$D_t$ is absent in the
definition of $E_i$, because, the combination $i[A_0,C_i]$ has been
absorbed to produce $D_t\Phi^j$ for both parts of $B_{ji}$.}. 

It is a good exercise to study the behaviour of eqs. (\ref{lorentz}) and
(\ref{A0}) under gauge transformation (\ref{tot-tra}). Since the action is
invariant under (\ref{tot-tra}), it is expected that the equations of
motion change covariantly. The left-hand side of (\ref{lorentz}) changes to
$U^\dagger D_tD_t\Phi U$ by (\ref{DTF}), and therefore we should find the
same change for the right-hand side. This is in fact the case, since
\bea
f(\Phi,t) &\goes& \tilde{f}(\tilde{\Phi},t)=
U^\dagger(\Phi,t) f(\Phi,t)U(\Phi,t),\nonumber\\
\frac{\delta f(\Phi,t)}{\delta \Phi^i} &\goes& 
\frac{\tilde{\delta}\tilde{f}(\tilde{\Phi},t)}{\delta \tilde{\Phi}^i}=
U^\dagger(\Phi,t) \frac{\delta f(\Phi,t)}{\delta
\Phi^i}U(\Phi,t),\nonumber\\
\frac{\partial f(\Phi,t)}{\partial t} &\goes&
\frac{\partial \tilde{f}(\tilde{\Phi},t)}{\partial t}=
U^\dagger(\Phi,t) \frac{\partial f(\Phi,t)}{\partial t} U(\Phi,t),
\eea
in which $\partial_i$ has been realized via its functional form,
$\delta/\delta\Phi^i$. In conclusion, the definitions (\ref{elec}) and
(\ref{magn}), lead to
\bea
E_i(\Phi,t) &\goes& \tilde{E_i}(\tilde{\Phi},t)=
U^\dagger(\Phi,t) E_i(\Phi,t)U(\Phi,t),\nonumber\\
B_{ji}(\Phi,t) &\goes& \tilde{B}_{ji}(\tilde{\Phi},t)=
U^\dagger(\Phi,t) B_{ji}(\Phi,t)U(\Phi,t),
\eea
a result consistent with the fact that $E_i$ and $B_{ji}$ are functionals
of $\Phi$. We thus see that, in spite of the absence of the usual
commutator term $i[A_\alpha,A_\beta]$ of non-Abelian gauge theories, in our
case the field strengths transform like non-Abelian ones. We recall that
these are all consequences of the matrix coordinates of D0-branes.  Finally
by the similar reason of vanishing of (\ref{vanishing}), both sides of
(\ref{A0}) transform identically. 

An equation of motion similar to (\ref{lorentz}) is considered in
\cite{fat241,fat021} as a part of similarities between the dynamics of
D0-branes and bound states of quarks--QCD strings
\cite{fat241,fat021,02414}. The point is that the center-of-mass dynamics
of D0-branes is not affected by the non-Abelian sector of the background,
\ie the center-of-mass is ``white" with respect to $SU(N)$ sector of
$U(N)$. The center-of-mass coordinates and momenta are defined by:
\bea 
\Phi^i_{c.m.}\equiv \frac{1}{N}
\tr \Phi^i,\;\;\;\;  P^i_{c.m.}\equiv \tr P^i_\Phi, 
\eea 
where we are using the convention $\tr {\bf 1}_N=N$. To specify the net
charge of a bound state, its dynamics should be studied in zero magnetic
and uniform electric fields, \ie$B_{ji}=0$ and $E_i(\Phi,t)=E_{0i}$
\footnote{In a non-Abelian gauge theory an uniform electric field can be
defined up to a gauge transformation, which is quite well for
identification of white (singlet) states.};  thus these fields are not
involved by $\Phi$ matrices, and contain just the $U(1)$ part. In other
words, under gauge transformations $E_{0i}$ and $B_{ji}=0$ transform to
$\tilde{E}_i(\Phi,t)=V^\dagger(\Phi,t)E_{0i}V(\Phi,t)=E_{0i}$ and
$\tilde{B}_{ji}= 0$. Thus the action (\ref{action}) yields the following
equation of motion: 
\bea
(Nm_0){\ddot{\Phi}}^i_{c.m.}=\mu_0N E^i_{0(1)},
\eea
in which the subscript (1) denotes the $U(1)$ electric field.  So the
center-of-mass only interacts with the $U(1)$ part of $U(N)$. From the
String Theory point of view, this observation is based on the simple fact
that the $SU(N)$ structure of D0-branes arises just from the internal
degrees of freedom inside the bound state. 

It will be interesting to mention the relation between the dynamics of
D0-branes in RR background, and the semi-classical equations of motion for
charged particles in Yang-Mills background. By semi-classical, as will be
more clear later, we mean treating the space-time motion of charged
particles classically, while describing the charge degrees of freedom,
calling them ``isotopic spin," quantum mechanically. The classical
mechanics of the charged particles is known by the original work of Wong
\cite{wong}, based on an appropriate limit of the equations of motion of
operators; see \cite{book} as a good review. The starting point is the
standard action of $U(N)$ gauge theory, accompanied with fermionic matter
in the fundamental representation (in units $c=1,\; \hbar=1$) 
\bea
S=\int d^dx \bigg(
-\bar\psi \gamma^\mu (\partial_\mu +i A_\mu)\psi-m\bar\psi \psi
-\frac{1}{4g^2}\tr F_{\mu\nu}F^{\mu\nu}
\bigg),\\
\mu, \nu=0,\cdots,d-1,\;\;\;A^\mu=A^\mu_aT^a,\;\;\;
F^{\mu\nu}=F^{\mu\nu}_aT^a.
\eea
In above $T^a$, $a=1,\cdots, N^2$, are $N\times N$ matrices as generators
of the $U(N)$ group, with the commutation relation $[T^a,T^b]=if^{ab}_c
T^c$; we assume sum over the lower and upper indices. Consequently, the
classical equation of motion for the charged particle are proposed as
\bea\label{wong1}
m\frac{d^2 \xi^\mu(\tau)}{d\tau^2}=
\bigg(F^{\mu\nu}_a(\xi)\tcl^a(\tau)\bigg)
\frac{d\xi_\nu(\tau)}{d\tau},
\eea
where $\xi^\mu(\tau)$ denote the world-line of the particle, parametrised
with $\tau$. In above, $\tcl^a(\tau)$ are numbers as the classical
analogues of the matrices $T^a$, with the canonical relation
$\{\tcl^a(\tau),\tcl^b(\tau)\}=f^{ab}_c\tcl^c(\tau)$, and the equation of
motion as
\bea\label{wong2}
\frac{d\tcl^a(\tau)}{d\tau}-\frac{d\xi_\mu}{d\tau}
f^{ab}_c A^\mu_b(\xi) \tcl^c(\tau) =0.
\eea
Thus, the particle is described by an internal vector $\tcl^a(\tau)$ as
well as its space-time coordinates $\xi^\mu(\tau)$. Also, by the equations
for $\tcl^a(\tau)$ one deduces that $d/d\tau(\tcl^a{\tcl}_a)=0$. Hence the
isotopic spin of the particle performs a precessional motion.

Now we try to sketch the relation. Let us first take
the simple case of one charged particle. One may define the covariant
derivative along the world-line by notion of world-line gauge field 
$A_\tau\equiv\dot\xi_\mu A^\mu_{\rm cl}(\xi,\tau)=\dot\xi_\mu A^\mu_a(\xi)
\tcl^a(\tau)$, as follows
\bea\label{cder}
D_\tau\equiv  \partial_\tau -\{ A_\tau,\;\}.
\eea
We notice that, from the world-line theory point of view, $A_\tau$ is 
a dynamical variable. Hence the equation of motion (\ref{wong2}) reduces
to $D_\tau \tcl^a(\tau)=0$. Then we define the variables as 
$X^{\mu a}(\tau)\equiv \xi^\mu(\tau) \tcl^a(\tau)$; for these variables
we have
\bea
D_\tau X^{\mu a}=\dot\xi^\mu(\tau) \tcl^a(\tau).
\eea 
Also by using (\ref{wong1}) we find that
\bea
mD_\tau D_\tau X^{\mu a }&=&m\ddot\xi^\mu(\tau) \tcl^a(\tau) =
F^{\mu\nu}_{\rm cl}(\xi,\tau)\dot\xi_\nu \tcl^a(\tau)\Rightarrow\nonumber\\
mD_\tau D_\tau X^{\mu a }&=& F^{\mu\nu}_{\rm cl}(\xi,\tau) D_\tau
X_\nu^a,
\eea
in which we are using the notation $F^{\mu\nu}_{\rm cl} (\xi,\tau)
=F^{\mu\nu}_a(\xi)\tcl^a(\tau)$.  The last equation for $X^a$ are
reminiscent of the D0-brane equation of motion in the RR 1-form
background, with this exception that here the field strength
$F^{\mu\nu}_{\rm cl}(\xi,\tau)$ depends on variables $\xi^\mu$. Here one
can define a map from fields $F^{\mu\nu}_{\rm cl}(\xi,\tau)$ to new fields
$\hat F^{\mu\nu}(X^a,\tau)$. The map defines the fields $\hat
F^{\mu\nu}(X^a,\tau)$ by the relation
\bea
F^{\mu\nu}_{\rm cl}(\xi,\tau)=F^{\mu\nu}_a(\xi)\tcl^a(\tau)\equiv \hat
F^{\mu\nu}(X^a,\tau).
\eea
Thus we have the equations: 
$mD_\tau D_\tau X^{\mu a}= \hat F^{\mu\nu}(X^a,\tau) D_\tau X_\nu^a$.
The map also may be defined at the level of gauge potentials 
$A^\mu_{\rm cl}(\xi,\tau)$ and $\hat A^\mu(X^a,\tau)$, in a Lagrangian or
Hamiltonian formulation of the problem.

The last equation we obtained have already the extra index $a$ on the
variables $X^{\mu a}$. One may consider $N^2$ copies of variables
$\xi^\mu_a (\tau)$, with common charge variables $\tcl^a(\tau)$. Since in
this case there are $N^2$ copies of variables, and so world-lines, the
definition of a unique world-line covariant derivative $D_\tau$ is not
possible. The best one can think about, as the case for coincident
D0-branes, is that the $N^2$ copies are nearly ``identified" in
space-time. Thus we assume $\xi_a (\tau)= \xi (\tau) +\delta\xi_a (\tau)$,
with $\delta\xi_a\ll \xi$. Actually the thing one needs is a unique
combination of $A_\tau=\dot\xi_{a\mu} A^\mu_{\rm cl}(\xi_a,\tau)$ as a
unique world-line gauge field. So to have a unique world-line gauge
field, and consequently common charge variables $\tcl^a(\tau)$, we have
the condition $\dot\delta\xi_{a\mu}A^\mu_a+\dot\xi_{a\mu}\delta\xi^\rho_a
\partial_\rho A^\mu_a=0$, for each $a$. Therefore we can define
\bea
D_\tau\equiv \partial_\tau -\{A_\tau,\;\},
\eea
with $A_\tau\equiv\dot\xi_\mu A^\mu_{\rm cl}(\xi,\tau)$. Thus all of the
charges can be taken equal, satisfying $D_\tau\tcl^a(\tau)=0$. Then we
define $X^\mu(\tau)=\xi^\mu_a(\tau) \tcl^a(\tau)$, and accordingly we have
\bea
mD_\tau D_\tau X^\mu&=& m\ddot\xi^\mu_a\tcl^a,\nonumber\\
mD_\tau D_\tau X^\mu&=& F^{\mu\nu}_{\rm cl}(\xi_a,\tau) 
\dot\xi_{a\nu}\tcl^a(\tau),\nonumber\\
mD_\tau D_\tau X^\mu&=& \bigg(F^{\mu\nu}_b(\xi_a)\tcl^b(\tau)\bigg) 
\dot\xi_{a\nu}\tcl^a(\tau),
\eea
where in the right-hand side the sums on $a$ and $b$ are recalled.
The dependence of field strength on $\xi_a$ prevents us to do the sums on
$a$ and $b$ independently, to get $\dot\xi_{a\nu}\tcl^a(\tau)=D_\tau
X_\nu$. Like the case for a single particle, we can
define the map between fields $F^{\mu\nu}_{\rm cl}(\xi_a,\tau)$ and 
$\hat F^{\mu\nu}(X,\tau)$ such that the expression in the right-hand side
appears as following
\bea
\bigg(F^{\mu\nu}_b(\xi_a)\tcl^b(\tau)\bigg)\dot\xi_{a\nu}\tcl^a(\tau)
\equiv 
\hat F^{\mu\nu}(X,\tau) D_\tau X_\nu.
\eea
The map can also be presented in components with group indices. Finally
one concludes with equations
\bea
mD_\tau D_\tau X^\mu= \hat F^{\mu\nu}(X,\tau) D_\tau X_\nu.
\eea

All the relations we had in above, rather than matrices, were about
numbers unfortunately! It is because that in Wong's theory also the
charge variables are treated with their classical analogues. It may be
possible to find a semi-classical version of the problem, assuming
space-time motion classically, while the isotopic spin $T^a$'s remain
matrix variables, as they should be as group generators; similar to the
situation we have for ordinary spin in Stern-Gerlach experiment. Then, a
relation between the semi-classical dynamics of charged particles and
D0-brane dynamics will appear very interesting. It remains for future
progresses to know more about both D0-brane dynamics and charged particle
dynamics in Yang-Mills background. 

All the above can be considered in non-relativistic limit, though
the case needs more study, may be defined by
\footnote{For the case concerning more than one massless particle, the
more systematic way can be going to the light-cone gauge to recover
non-relativistic dynamics in the transverse directions; see Appen. of
\cite{fat021}.}
\bea
D_\tau X^0 \simeq 1, \;\;\;\; \dot\xi^i_a\ll 1.
\eea
In this limit we have $A_\tau(t)=A_{0a}(t,\xi^i)\tcl^a(t)+O(\dot\xi^i)$,
which means that the world-line gauge field equals effectively to zero
component of gauge field, which for very small velocities may be assumed
to be function of $t$ only.

In the above we saw the important role of a map between the Yang-Mills
fields, and fields which depend on $X^\mu$ variables. In \cite{9908142} a
map between field configurations of non-commutative and ordinary gauge
theories is introduced, which preserves the gauge equivalence relation. It
is emphasized that the map is not an isomorphism between the gauge groups.
It will be interesting to study the properties of the map between
non-Abelian gauge theory and gauge theory associated with matrix
coordinates of D0-branes; on one side the quantum theory of matrix fields,
and on the other side the quantum mechanics of matrix coordinates. Since
in this case we have matrices on both sides, it may be possible to find an
isomorphism between all objects involving in the two theories, \ie
dynamical variables and transformation parameters.

We have seen how the $U(1)$ symmetry of the 1-form RR field
$C^{(1)}_\mu(x^\nu)$ can show a $U(N)$ structure inside the bound states.
The $U(N)$ structure in bound states is a consequence of the fact that the
correct dynamical variables of the bound states are $N\times N$ hermitian
matrices, rather than numbers. From the String Theory point of view, this
is possible by taking into account degrees of freedom corresponding to $N$
copies of the $U(1)$ structure, together with $N^2-N$ additional ones
coming from the dynamics of strings stretched between charged particles.
Here each D0-brane carry $1/N$ fraction of the bound state total charge.
This is an example of a mechanism that how small (fractional) charges of
an Abelian symmetry can form a bound state with an internal non-Abelian
symmetry; a mechanism which may be called ``non-Abelian from fractional
Abelian charges." 

{\bf Acknowledgement:} The comments on manuscript by S. Parvizi, M.M. 
Sheikh-Jabbari and specially A. Sagnotti are deeply acknowledged. Also the
author is grateful to M. Khorrami and A. Shariati at IASBS Zanjan, for
useful discussions on the dynamics of charged particles. 

\end{document}